\documentclass[preprint]{elsarticle}

\usepackage{graphicx} 
\usepackage{amsmath,amssymb}
\usepackage{color}
\usepackage{multirow}
\usepackage{booktabs}
\usepackage{url}
\usepackage{lineno}  
\usepackage{tipa} 
\usepackage{subcaption} 
\usepackage{appendix}

\usepackage{ulem} 

\journal{Speech Communication}

\biboptions{authoryear}

\begin{document}

\begin{frontmatter}
\title{Disentangling peripheral hearing loss  \\ from central and cognitive effects  \\on speech intelligibility in older adults
\tnoteref{mytitlenote}}



\author[WUc,WUg,UEC]{Toshio Irino}
\ead{irino@wakayama-u.ac.jp}

\author[WUg]{Ayako Yamamoto}
\ead{yamamoto.ayako@g.wakayama-u.jp}

\author[WUg]{Fuki Miyazaki}
\ead{miyazaki.fuki@g.wakayama-u.jp}

\address[WUc]{Center for Innovative and Joint Research, Wakayama University, 930 Sakaedani, Wakayama,  640--8510, Japan}

\address[WUg]{Graduate School of Systems Engineering, Wakayama University, 930 Sakaedani, Wakayama, 640--8510, Japan}

\address[UEC]{ School of Informatics and Engineering, the University of Electro-Communications,	1-5-1 Chofugaoka, Chofu, Tokyo, 182-8585, Japan}

\begin{abstract}
Age-related hearing loss (HL) reduces speech intelligibility (SI) in older adults (OAs). However, deficits in central and cognitive processing also substantially impact SI. Understanding these contributions is essential for explaining individual differences and developing effective assistive hearing strategies. This study presents a framework that distinguishes peripheral HL from central and cognitive influences on SI. This framework uses the Wakayama University Hearing Impairment Simulator (WHIS), and the Gammachirp Envelope Similarity Index (GESI), an objective measure of intelligibility.  First, speech-in-noise tests were conducted with young, normal-hearing listeners (YNHs) using WHIS to simulate the audiogram of a target OA. The target OA achieved SI scores comparable to or higher than those of YNHs with simulated HL, suggesting contributions beyond peripheral hearing function. Then, GESI was used to predict SI scores for YNHs and OAs across different hearing levels. The prediction accuracy was comparable for both groups. Interestingly, many OAs' subjective SI scores were higher than those predicted using parameters derived from YNHs' experiments. This finding is inconsistent with previous research indicating that speech perception ability declines with age. This issue will be discussed. There was no significant correlation between the average hearing levels and the residual differences between the subjective and predicted SI scores.  This suggests that GESI effectively absorbed the effects of peripheral HL. Thus, the proposed framework may facilitate systematic examination and comparison of central and cognitive factors beyond peripheral HL among individual YNHs and OAs with and without HL. 



\end{abstract}



\end{frontmatter}



\section{Introduction}
\label{sec:Introduction}

Age-related hearing loss (HL) impairs the ability to communicate through speech
-- the foundation of human interaction -- and consequently diminishes quality of life~\citep{dalton2003impact}. Hearing aids are one of the most effective interventions for hearing difficulties and primarily aim to compensate for peripheral HL, achieving notable success~\citep{dillon2012hearing}.
However, current devices rarely consider many factors related to central and supra-threshold auditory and cognitive processing~\citep{schoof2014role,fullgrabe2015age}. Supra-threshold auditory processing (e.g., ~\citealt{summers2013suprathreshold,  kortlang2016suprathreshold}) involves temporal envelope modulation, temporal fine structure, spectral modulation, loudness, and binaural effects. 
Cognitive factors include access to the mental lexicon, listening effort, working memory capacity, processing speed, and attention~(e.g., ~\citealt{dubno2000use, gordon2009aging, holt2022speech}). These factors vary substantially across individuals and listening situations.
When designing new hearing assistive devices for older adults (OAs), it is important to first evaluate the effects of these factors separately from the effects of peripheral HL.

 

Previous studies have examined the role of central/cognitive processing in speech perception. For example, \cite{schoof2014role} and \cite{fullgrabe2015age} compared auditory and cognitive processing between young normal-hearing (YNH) and older normal-hearing (ONH) listeners using psychoacoustic and cognitive tests, including speech intelligibility (SI) in noise. They concluded that declines in speech perception in ONH are partly due to central/cognitive factors. However, because the hearing levels were matched, the study could not address the relative contributions for OAs with HL.
Similarly, \cite{moore2014relation} investigated the relationship between hearing level, speech reception threshold (SRT), and cognition in listeners aged 40–69, but did not include YNH. \cite{kocabay2022speech} examined the effects of hearing level, temporal processing, and cognitive processing on SI for listeners aged 18–59, yet excluded OAs. \cite{bugannim2025speech} analyzed age effects on SRT, hearing level, and cognition across ages 7–90, but did not determine individual contributions.
Overall, previous studies support the view that central/cognitive processing tends to decline with age~\citep{kortlang2016suprathreshold}. However, the individual characteristics of OAs -- essential for the development of personalized hearing assistive technologies -- have often been underexplored. 

It is therefore desirable to identify which functions are degraded and to what extent in each OA relative to YNH listeners.
A major challenge for direct comparison with YNH listeners lies in two factors: the large variability in peripheral HL among OAs and the fact that
the effects of central/cognitive processing may sometimes be less dominant than peripheral HL.
To overcome this, we aim to provide a method that isolates the impact of central/cognitive processing by removing the influence of peripheral HL.
This approach will enable a more precise understanding of age-related deficits beyond peripheral HL, paving the way for hearing aid designs that incorporate cognitive and supra-threshold auditory factors.

This study addresses this issue by proposing a method that uses a hearing loss simulator (HLS) and an objective intelligibility measure (OIM) developed by the authors.
The use of an HLS is one approach: SI experiments with YNH listeners are conducted using speech signals degraded to replicate the HL of a target OA. This allows a direct comparison of SI between the OA and YNH listeners, where differences in SI scores may indicate the effects of central/cognitive processing in the target OA.
This approach has been reported by \cite{fontan2017automatic} and \cite{fullgrabe2022immediate} using the HL simulator (HLS) developed by the Cambridge Hearing Group (CamHLS; \citealt{nejime1997simulation}), which is widely used for SI experiments. We have also proposed WHIS (Wakayama-University Hearing Impairment Simulator; \citealp{irino2023hearing}), which provides better sound quality and more precise absolute threshold reproduction than CamHLS~\citep{irino2024signal, deutsch2026evaluation}. 
SI experiments using WHIS have been reported in \cite{irino2022speech} and \cite{yamamoto2023gesi}, where the experiments were conducted in both laboratory and remote environments. However, these studies did not include a direct comparison between NH and HL listeners.
In this study, we conducted an SI experiment with YNH listeners using WHIS, which simulates the HL of a specific OA identified in a previous OA study~\citep{yamamoto2025predicting}.
However, this method only allows us to examine the characteristics of one OA at a time. We would like to investigate various OAs effectively.

We propose a new approach of using the GESI (Gammachirp Envelope Similarity Index;  \citealt{irino2022speech, yamamoto2025predicting}) to predict SI scores for OAs, with parameters determined from YNH listeners. 
\cite{yamamoto2025predicting} demonstrated that GESI predicts SI scores for words more accurately than the conventional metric HASPIw2~\citep{kates2023extending}. 
GESI is a simple SI prediction model and is based on the gammachirp filterbank (GCFB; ~\citealp{irino2006dynamic,irino2023hearing}), a modulation filterbank, and a similarity measure, as briefly described in \ref{sec:Appendix_GESI}.  Unlike other OIMs with neural networks (NNs), all the parameters were explicitly defined. The SI metric is calculated within the peripheral process and part of the central process and converted into an SI score using a simple sigmoid function. Therefore, anything that cannot be predicted may be considered a factor not modeled in GESI. It could be used to separate the effects of peripheral and central/cognitive factors, as described below.



In the following sections, we first present the results of the SI experiments with YNH listeners using WHIS. Second, We demonstrate that GESI can predict SI reasonably well in both YNH and OA listeners, regardless of peripheral HL. Then, we present the prediction of the SI scores in the previous OA study using GESI with the parameters obtained from the YNH experiment.

\color{black}

\section{Experiment in YNH with WHIS} 
\label{sec:Exp}
The SI experiment was conducted when YNHs listened to degraded speech synthesized to reflect the HL of a specific OA using the HL simulator, WHIS~\citep{irino2023hearing}. 
We compare the results of the target OA and YNHs. 
This experiment follows a design similar to that used in the previous OA study~\citep{yamamoto2025predicting}.

\subsection{Participants and characteristics } 
\label{sec:ExpYNH_Listener}

Fourteen young adults, ranging in age from 18 to 21, participated in the experiment. Their native language is Japanese. They are all unfamiliar with this type of SI experiment, which is designed to prevent listeners from recalling and responding words.
The hearing levels obtained by pure-tone audiometry are 20 dB or less, as shown in Fig.~\ref{fig:Audiogram_YNH} (\ref{sec:Appendix_GESI}). 
In contrast, Fig.~\ref{fig:Audiogram_OA} shows the hearing levels of fifteen OAs, ranging in age from 62 to 81, who participated in the previous OA study. Clearly, the hearing levels of the YNHs are better than those of the OAs.
We calculated  the four-frequency pure-tone average (PTA4) for each ear, which is the average hearing level from 500 to 4,000 Hz. The ear with the lower value was considered the better ear.

The temporal modulation transfer function (TMTF) was also measured using the two-point method (\citealt{morimoto2019twopoint} and Appendix B of \citealt{yamamoto2025predicting}) for rapid measurement.
Figure~\ref{fig:TMTF_YNH} shows the results.
This method approximates the TMTF as a first-order low-pass filter (LPF) as shown in this figure.
There are two parameters: the peak sensitivity at low modulation frequencies ($L_{ps}$) in dB and the cutoff frequency of the LPF ($F_c$).

\subsection{Speech materials}
\label{sec:Exp_SpeechMaterial}

We used the speech words pronounced by a male speaker (``mis'') from the minimum familiarity rank data in the Japanese four-mora word dataset FW07~\citep{sakamoto2006new}. This dataset contains 20 lists, each with 20 words. The FW07 database is a subset of the well-designed FW03 database for controlling word familiarity~\citep{amano2009development}. The speech sounds were processed as follows.

Babble noise was used as background noise to control the signal-to-noise ratio (SNR). This noise was produced by combining and overlapping speech sounds from the FW03 database in order to match the long-term spectrogram with the target sounds. 
To simulate a realistic listening environment, room reverberation was applied to the speech sounds. Room impulse responses (RIRs) obtained from the Aachen database~\citep{jeub2009binaural}. The target speech was convolved with the RIR  at a distance of 2 m in an office room. The babble noise was convolved with the RIRs at 1~m and 3~m separately and then added together to create a more diffuse sound.  The SNR was set between -6 dB and 12 dB, increasing by 6~dB. This is referred to as the unprocessed condition (hereafter ``Unpro'') because no speech enhancement algorithm was applied.

We also prepared word sounds processed by speech enhancement using an ideal ratio masker (IRM) to evaluate its effectiveness on SI for OAs. The ``Unpro'' sounds of the other words were processed with the IRM. This condition is referred to as ``IRM.'' 
The ``Unpro'' and ``IRM'' sets of sounds are the same as those used in the previous OA study~\citep{yamamoto2025predicting}. 
The use of IRM, as described in Appendix C of ~\citealt{yamamoto2025predicting}, has been proposed to provide oracle data for training the DNN-based speech enhancement algorithms~\citep{wang2014training}. 
Therefore, the SI scores for the trained DNNs are expected to fall between the  ``Unpro''  and ``IRM''  scores.


In the current study, these sets were processed with WHIS using the hearing levels of a target OA for HL simulation. The target listener and the rationale for its selection will be explained below.
These sets are referred to as ``Unpro-WHIS'' and ``IRM-WHIS.''
The fifth set was produced from clean, non-reverberant speech sounds, referred to as ``Dry-Unpro,'' that were extracted directly from FW07. 
The sounds were mixed with the same babble noise to produce SNRs of -6, 0, 6, and 12 dB.
This condition was introduced to examine the effect of reverberation on subjective SI scores and the objective prediction by GESI. 
Thus, there were five sound processing conditions and four SNR conditions. Each list contained 20 words.
The total number of words was 400 (= 20 $\times$ 5 $\times$ 4).

\begin{figure}[t]
    \centering
        \begin{minipage}{0.32\textwidth}
        \centering
        \includegraphics[width=1.1\columnwidth]{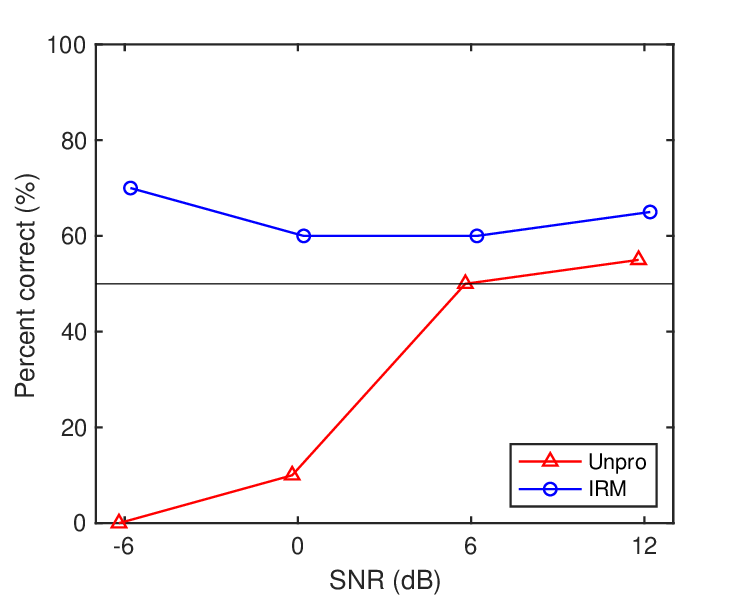} 
        \subcaption{SI score}
        \label{fig:SI_MNM}
    \end{minipage}
    \begin{minipage}{0.32\textwidth}
        \centering
        \includegraphics[width=1.1\columnwidth]{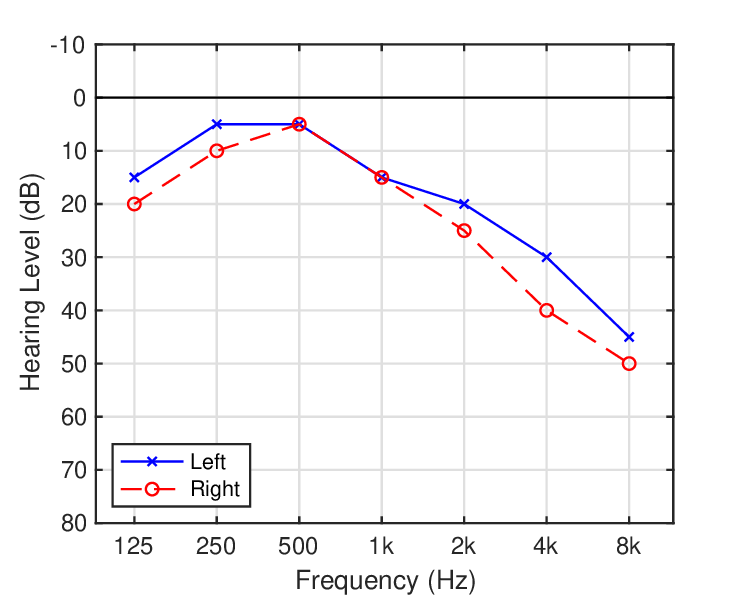}
        \subcaption{Audiogram}
        \label{fig:Audiogram_MNM}
    \end{minipage}
    \begin{minipage}{0.32\textwidth}
        \centering
        \includegraphics[width=1.1\columnwidth]{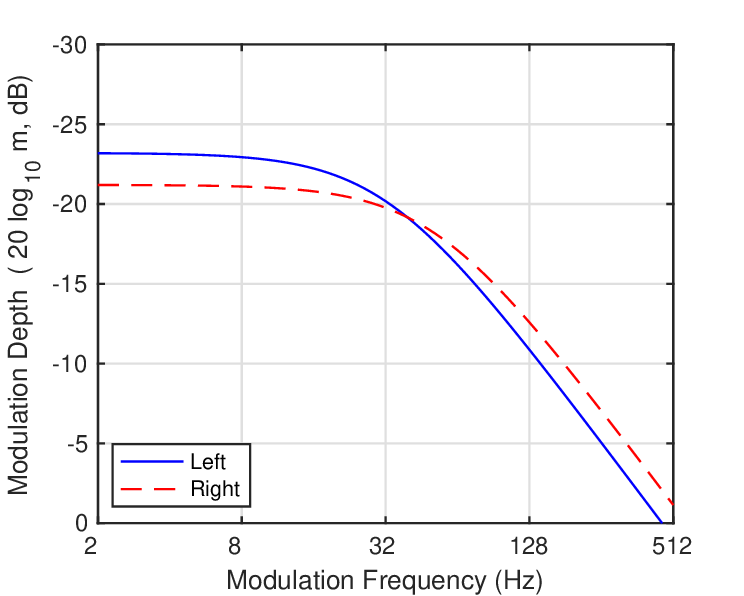} 
        \subcaption{TMTF}
        \label{fig:TMTF_MNM}
    \end{minipage}

\caption{ SI score, audiogram, and TMTF of $\rm OA^{\#7}$, the target OA listener for HL simulation. Data were obtained from the previous OA study~\citep{yamamoto2025predicting}. }
\label{fig:ExpEldIRM23_MNM}
\end{figure}


\subsection{HL simulation}
\label{sec:HearingLossSimulation}

The HL simulator WHIS~\citep{irino2023hearing} was used in this study. WHIS first analyzes input sounds using the gammachirp filterbank (GCFB; ~\citealp{irino2006dynamic, irino2023hearing}) to produce an auditory spectrogram -- a sequence of excitation patterns (EPs) referred to as an EPgram. In the GCFB analysis, the total hearing loss ($HL_{total}$) shown on the audiogram is assumed to be the sum of the level-dependent active HL ($HL_{act}$) and the level-independent passive HL ($HL_{pas}$) on a dB scale, in accordance with the loudness model proposed by \cite{moore1997model}. The total HL was set to match the hearing levels of the target OA listener. This procedure enables the precise reproduction of absolute thresholds, as demonstrated in psychological experiments~\citep{deutsch2026evaluation}. 

The degree of active hearing loss ($HL_{act}$) can be controlled by a parameter $\alpha$, which represents the health of the cochlear compressive input–output function (for details, see Appendix A in \citealt{yamamoto2025predicting}). In this study, $\alpha$ was set to 0.5, corresponding to an intermediate level between no dysfunction ($\alpha = 1$) and complete damage ($\alpha = 0$). For the HL simulation, gain reduction was calculated using the EPgrams of a target OA and a typical NH listener. HL sounds were then synthesized using the direct time-varying filter (DTVF) method, which produces less perceptual distortion than CamHLS~\citep{irino2024signal}.

\subsection{Simulation target OA}
\label{sec:ExpYNH_TargetOA}

In the previous OA study, 15 OA participants were included, and the seventh participant ($\rm OA^{\#7}$) was selected as the target for WHIS simulation.
This is because the OA had an audiogram consistent with typical age-related HL and demonstrated a thorough understanding of the experimental procedure.
Figure~\ref{fig:ExpEldIRM23_MNM} shows the SI score, audiogram, and TMTF.
The PTA4 for the better ear was 17.5 dB. The TMTF values were $L_{ps}$ = -23.2 dB and $F_{c}$ = 51~Hz, which were similar to the average YNH values. Small differences were observed between the left and right ears for both HL and TMTF.

\subsection{Procedure}
\label{sec:ExpYNH_Procedure}

There were five sound processing conditions: ``Unpro,'' ``IRM,'' ``Unpro-WHIS,'' ``IRM-WHIS,'' and ``Dry-Unpro.''
The sounds were presented through a web-based GUI system~\citep{yamamoto2021comparison}.
Four hundred words were presented in 40 sessions, with 10 words in each session. Each word was played, followed by a four-second response period before the next word was played. The percentage of correct words was recorded for each of the five sound processing conditions and the four SNR conditions.

Each participant listened to a different set of words.
The participants were seated in a sound-attenuated room (YAMAHA AVITECS) with a background noise level of approximately 26.2~dB in $L_{\rm Aeq}$.
The sounds were presented diotically through a DA-converter (SONY, Walkman 2018 model NW-A55) connected to a computer (Apple, Mac mini) via headphones (SONY, MDR-1AM2). 
The sounds of the conditions ``Unpro,'' ``IRM,'' and ``Dry-Unpro'' were presented at a sound pressure level (SPL) of 63~dB $L_{\rm eq}$.
This is the same level as the calibration tone measured with an artificial ear (Br\"{u}el \& Kj\ae r, Type~4153), a microphone (Br\"{u}el \& Kj\ae r, Type~4192), and a sound level meter (Br\"{u}el \& Kj\ae r, Type~2250-L).

Due to the hearing loss simulation with WHIS, the sounds in "Unpro-WHIS" and "IRM-WHIS" have lower SPLs than the other three conditions.  
These sounds were assigned to different sessions than the 63-dB SPL sounds.
Moreover, before the words were presented in each session, a message in Japanese was played that said, ``The audio will play at this volume,'' allowing listeners to anticipate the loudness.
When sounds with different SPLs are played randomly, listeners may have difficulty concentrating on and recognizing quiet sounds because they are bracing for loud ones. This procedure eliminated that possibility.

Experimental details were explained with documentation, and informed consent was obtained in advance. The experiment was approved by the Wakayama University Ethics Committee (reference numbers~2015-3, Rei01-01-4J, and Rei02-02-1J).


\begin{figure}[t]
    \centering
    \begin{minipage}{0.49\textwidth}
        \centering
        \includegraphics[width=\columnwidth]{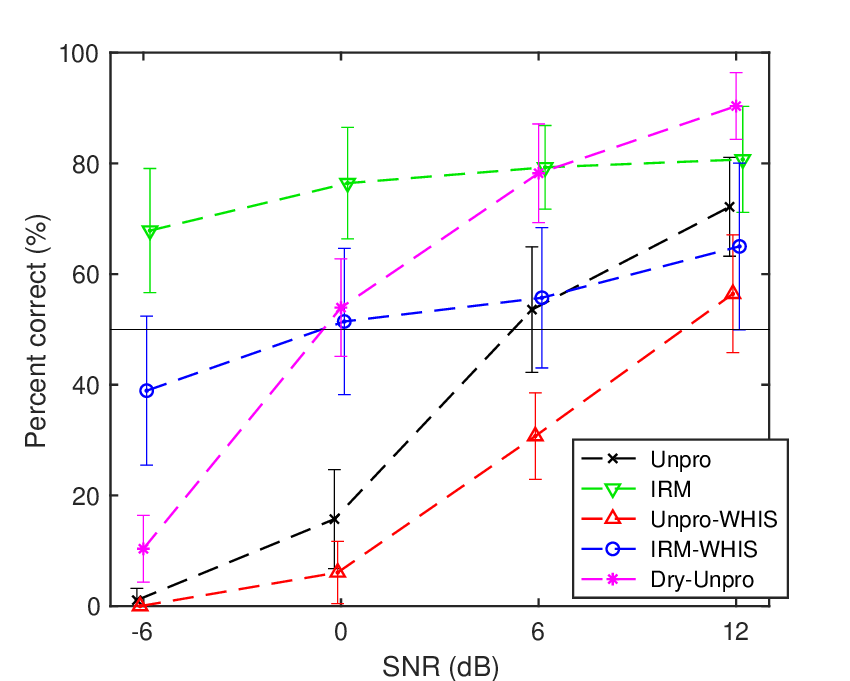}
        \subcaption{YNH experiment}
        \label{fig:YNH_SbjSI_AllCondPrcs}
    \end{minipage}
    \begin{minipage}{0.49\textwidth}
        \centering
        \includegraphics[width=\columnwidth]{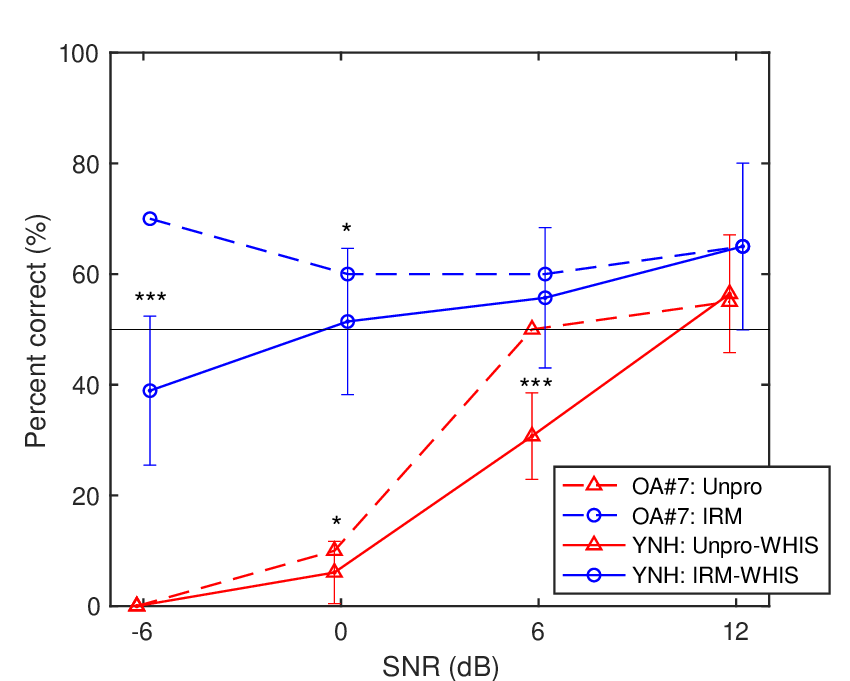} 
        \subcaption{ Comparison of $\rm OA^{\#7}$ and YNH with WHIS}
        \label{fig:CmprOA7vsYNH-WHIS}
    \end{minipage}

\caption{Subjective SI scores. (a)  Mean and standard deviation of 14 YNHs for all five sound processing conditions. 
(b) ``Unpro'' and ``IRM'' of $\rm OA^{\#7}$ (dashed line; replot of Fig.~\ref{fig:SI_MNM}) and ``Unpro-WHIS'' and ``IRM-WHIS'' of YNH (solid  line with error bar; extracted from Fig.~\ref{fig:YNH_SbjSI_AllCondPrcs})  The results of the $t$-tests are shown as: ***: $p < 0.001$ and *: $p<0.05$.   }
\label{fig:ExpWHISmnm24}
\end{figure}

\subsection{Result}
\label{sec:ExpYNH_Result}
The SI score was defined as the word correct rate (\%) as in the OA experiment. Figure~\ref{fig:YNH_SbjSI_AllCondPrcs}
shows the mean and standard deviation of the SI scores for the YNH listeners across the five sound processing conditions.
The score for ``IRM'' is the highest at low SNRs. The score for ``Dry-Unpro'' is the highest at an SNR of 12 dB. When WHIS is applied, the SI scores drop by 10-30 percentage points, as shown by the lines labeled ``IRM-WHIS'' and ``Unpro-WHIS.''
Clearly, there is a large variation in SI scores among YNH listeners. 
The results are unexpected if we assume that cognitive differences among individuals would be smaller. However, such variation has also been reported in experiments using HINT with YNH listeners~\citep{vermiglio2022relationship}.

Our main concern is the difference in the SI scores between 
$\rm OA^{\#7}$ and the YNH listeners with WHIS.
Figure~\ref{fig:CmprOA7vsYNH-WHIS} shows the comparison between them.
The individual YNH scores, which are the sources of variation, can be seen in Fig.~\ref{fig:YNH_PredSI_IndivLstn}, described later. 

The SI score of $\rm OA^{\#7}$ at 12~dB, where speech sounds can easily be recognized, is equivalent to the average YNH score. It is suggested that WHIS-based signal processing effectively simulates age-related peripheral HL. This is also the case regardless of the SNR. WHIS performs hearing loss simulation based on the inverse process of dynamic compressive auditory filtering with HL for any input signal~\citep{irino2023hearing}. There is no distinction between the target speech and the background babble noise. Therefore, the output sounds are processed equally at any SNR. 

The SI scores of $\rm OA^{\#7}$ are generally higher than  the average YNH scores at lower SNRs. Notably, the score of ``IRM'' of $\rm OA^{\#7}$ at -6~dB is much higher than that of YNH. Therefore, we performed a $t$-test on the individual SNRs and processing conditions to determine if there were any statistically significant differences in subjective SI scores between $\rm OA^{\#7}$ and YNH listeners.
The results were presented with the asterisks above the error bars. Statistically significant differences were observed at SNRs of 0~dB ($p< 0.05$) and 6~dB  ($p < 0.001$) in ``Unpro'' and at -6~dB ($p < 0.001$) and 0~dB ($p< 0.05$) in ``IRM.''
There were no differences in the other conditions. The SI scores of $\rm OA^{\#7}$ were better than those of the YNH listeners at conditions with an SNR of less than 12 ~dB. 

\subsection{Potential explanations}
\label{sec:ExpYNH_Explanation}

The results could be explained in several ways.

\begin{enumerate}

     \item There were artifacts or distortions due to the WHIS processing, particularly at low SNRs. However, this explanation is unlikely because WHIS processing can be performed uniformly regardless of SNR, as described above. Furthermore, WHIS offers superior sound quality compared to CamHLS~\citep{irino2024signal}. Moreover, WHIS offers more precise absolute threshold reproduction than other HLSs~\cite{deutsch2026evaluation}. Since CamHLS has been widely used and has produced numerous findings for SI experiments (e.g., ~\citealp{baer1993effects, baer1994effects, stone1999tolerable}), WHIS could be utilized more effectively.
   
    \item The stimulus words for $\rm OA^{\#7}$ were easier, particularly under at the conditions with $p< 0.001$. The 20 words in each list of stimulus sounds were randomly assigned to each condition. It is known that the difficulty level varies slightly from list to list~\citep{sakamoto2006new}. However, the standard deviation in the SI scores in FW03 (i.e. source database of FW07) was reported to be less than 10 percentage points~\cite{amano2009development}, and thus the observed differences are larger than that level. Additionally, significant differences were observed at 0 dB in both ``Unpro'' and ``IRM.''  It is unlikely that a separate easy list can be assigned to each of these conditions. Therefore, this alone is not considered sufficient to explain the results.

    \item For YNH listeners participating in SI experiments for the first time, the simulated HL sounds by WHIS may be unfamiliar and more difficult to recognize. However, if such speech sounds were generally more difficult to hear, this effect would occur regardless of the SNR; it would not be specific to an SNR of -6 dB. In other laboratory SI experiments using WHIS~\citep{irino2022speech, yamamoto2023gesi}, SI has been found to remain consistent across all SNR conditions when the SPL is reduced by 20 dB. This is because the speech level has a sufficient margin above the absolute threshold. The decrease in SI under the simulated HL condition is thought to be due to the SPL in the high-frequency region falling below the AT. This is similar to what occurs in people with OA and HL. Additionally, while not an SI experiment, research on perceiving speech emotion has shown that adding simulated HL virtually does not affect emotion perception~\citep{irino2024effects}. These findings suggest that familiarity with simulated HL sounds has a minimal influence.
     
    \item Cognitive abilities of $\rm OA^{\#7}$ surpass those of YNH listeners. This advantage may be realized when the effective SNR is high.  Since WHIS can only simulate peripheral HL, the difference in SI scores between $\rm OA^{\#7}$ and YNH listeners could be attributed to central or cognitive processing. Cognitive functions can adapt to the gradual progression of age-related HL.  These abilities can develop naturally through everyday conversations. 
    We will revisit this issue in Section~\ref{sec:Discussion_Improvement}.

\end{enumerate}


The fourth possibility seems to be the most likely at this point, although it contradicts the conventional findings that OA's abilities decline compared to those of YNH.
However, we cannot draw a conclusion based on results from just one OA.
It is also necessary to examine the characteristics of the other OAs.
However, repeating similar experiments for each OA is impractical. This would require generating simulated HL sounds for each of the remaining 14 OAs and recruiting 14 new YNHs for each OA. In the following sections, we will present an alternative method for analyzing the characteristics of the other OAs.


\section{Prediction of YNH's SI score using GESI}
\label{sec:PredYNH_usingGESI}

The aim of this study is to examine the extent to which the SI score can be explained by peripheral HL alone, and to distinguish its effects from those of subsequent central/cognitive processing.
For this purpose, we use an OIM called GESI~\citep{irino2022speech,yamamoto2025predicting}, as briefly described in \ref{sec:Appendix_GESI}. 
GESI involves GCFB, a modulation filter bank and a similarity calculation up to the derivation of the metric. Therefore this metric only reflects peripheral and some central processes, not cognitive processing. While the sigmoid function may reflect factors that could not be modeled in GESI, two parameters cannot reflect most complex cognitive factors. This, in turn, contributes to separating the effects of
peripheral and central/cognitive processing as described below.

In this section, we will examine whether GESI can predict SI scores for YNHs within the same framework as the previous OA study~\citep{yamamoto2025predicting}, despite the different hearing levels of participants. This provides the background for the OA analysis performed in the following sections.

\color{black}

\subsection{Procedure}
\label{sec:PredYNH_Procedure}
The GESI parameters were set to reflect the characteristics of the YNHs. The hearing levels of the better ears, as shown in Fig.~\ref{fig:Audiogram_YNH}, were individually set to the GCFB in GESI. The compression health parameter, $\alpha$, was set to one for all listeners, indicating that the cochlear input-output function is functioning properly. The individual TMTFs shown in Fig.~\ref{fig:TMTF_YNH} were also introduced, though their effect may be limited, as demonstrated in the previous OA study. The metric parameters, $\rho$ in Eq.~\ref{eq:similarity} and $\eta$ in Eq.~\ref{eq:w_i_Ef}, were set to 0.52 and 0.7, respectively.  These values were determined based on the previous OA study and preliminary fittings to YNHs' SI scores shown in Fig.~\ref{fig:YNH_SbjSI_AllCondPrcs}.  $I_{max}$ in Eq.~\ref{eq:sigmoid} was set to 85\%, which is the maximum SI score for FW07.
This procedure was nearly identical to that used in the previous OA study, except for the parameter values. 

The parameters $a$ and $b$ were estimated from the SI scores of the ``Unpro'' of the first five YNHs, $\rm YNH^{\#1}$ to $\rm YNH^{\#5}$, in order of participation. Five individuals were selected to allow for comparison with the previous OA study. Therefore, the following results reflect the average characteristics of the five YNHs.   For convenience, these values are referred to as ``$a_{_{YNH}}$'' and ``$b_{_{YNH}}$.''

%
%

\begin{figure}[t]
    \centering
    \begin{minipage}{0.49\textwidth}
        \centering
        \includegraphics[width=1\columnwidth]{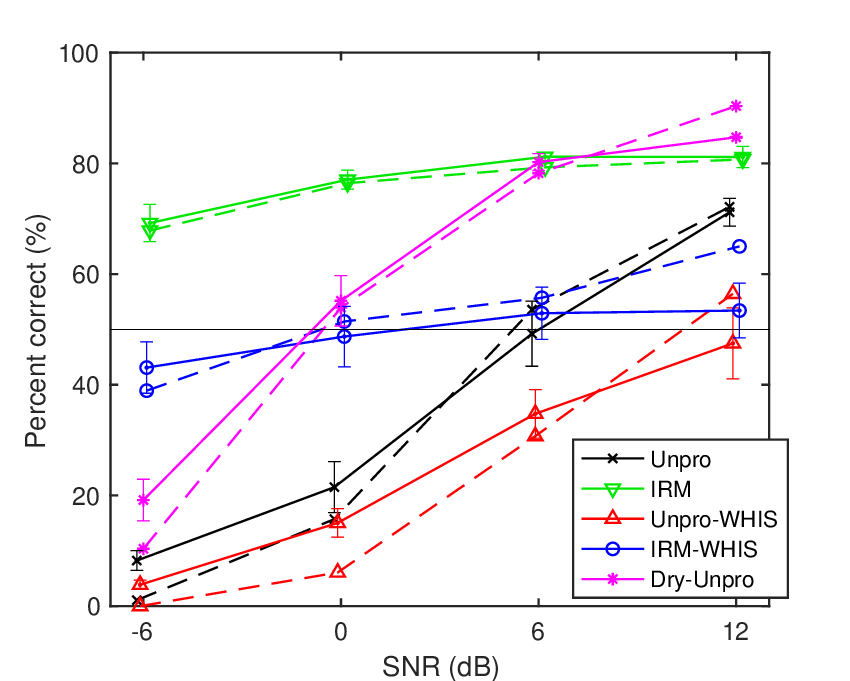}
      \subcaption{SI score }
        \label{fig:YNH_PredSI_AllCondPrcs}
    \end{minipage}
    \begin{minipage}{0.49\textwidth}
        \centering
        \includegraphics[width=1\columnwidth]{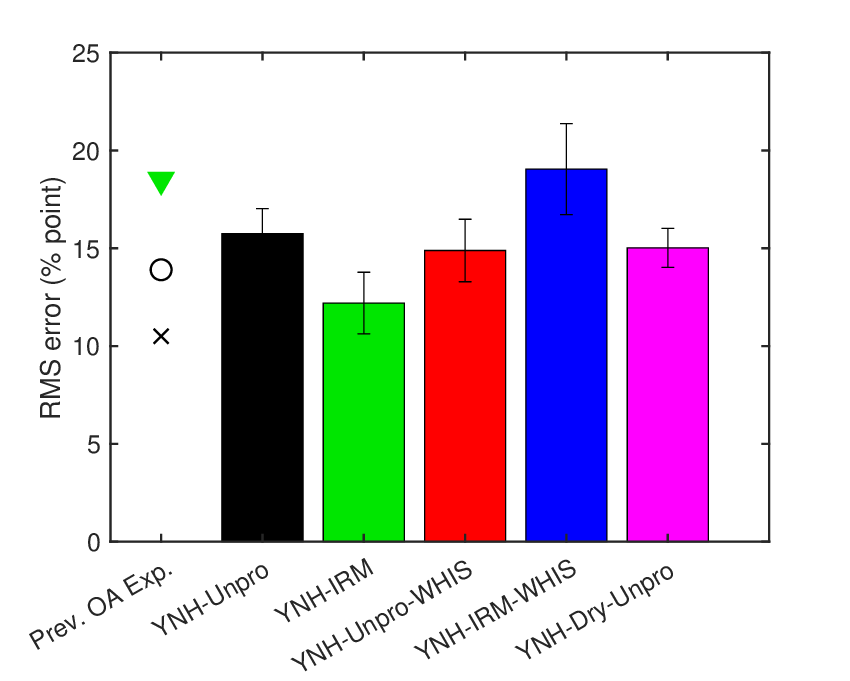} 
        \subcaption{RMS error}
        \label{fig:BarRMSE_YNH-WHIS}
    \end{minipage}
    \caption{Results of the SI prediction. (a) The predicted SI scores (solid line with error bars) are shown alongside the mean SI scores of YNHs (dashed line; extracted from Fig. ~\ref{fig:YNH_SbjSI_AllCondPrcs}). The error bar represents the standard deviation across the mean SI scores of individual listeners.
    (b) Mean RMSE (bar) and 95\% confidence interval (CI; error bar) for the five processing conditions. The left markers show the mean RMSEs reported in Fig. 5a of ~\cite{yamamoto2025speech}. ``Unpro (closed)'' is represented by the ``$\times$,'' ``Unpro (open)'' by the  ``$\circ$,'' and ``IRM'' by the ``$\blacktriangledown$.'' }
    
\end{figure}

\subsection{Prediction result}
\label{sec:PredYNH_Result}
Figure~\ref{fig:YNH_PredSI_AllCondPrcs} shows the predicted SI scores (solid lines with error bars) and the average subjective SI scores (dashed lines) for the five sound processing conditions. Overall, the two are in good agreement. Therefore, the GESI prediction accurately reflects the average scores.
The standard deviation in the prediction is quite small. This is likely due to the small variation in hearing levels among the YNH listeners, as shown in Fig.~\ref{fig:Audiogram_YNH}. In contrast, the variation in the subjective SI scores is much greater, as shown in Fig.~\ref{fig:YNH_SbjSI_AllCondPrcs}.


To evaluate prediction accuracy, the root-mean-square error (RMSE) was calculated for each participant by comparing their subjective SI scores with the predicted scores across different words and SNRs. The five bars on the left side of Fig.~\ref{fig:BarRMSE_YNH-WHIS} show the mean RMSE and 95\% confidence interval (CI) for all participants. 

The left markers show the mean RMSEs reported in Fig. 5a of in~\cite{yamamoto2025predicting}.
The mean values (95\% CIs) were  10.5 ($\pm 0.73$) for ``Unpro (closed)'' (black cross),   13.9 ($\pm 1.0$) for ``Unpro (open)'' (black circle), and 18.0 ($\pm 0.40$) for `IRM'' (green asterisk).
These values were derived from ten repeated estimations using different OAs to determine the sigmoid parameters. As these differ from the current one-time estimation, only the mean values are plotted here. Note that all the CI values were 1.0 or less and much less than the mean values. This indicates that the variability across repeated estimations is sufficiently small. This observation will be used in Section~\ref{sec:Discussion_Potential}.

It is clear that ``Unpro (closed)'' and ``Unpro (open)'' in the previous OA study are less than the mean value of ``YNH-Unpro'' (black bar; a mix of open and closed predictions). In contrast, ``IRM'' in the previous OA study is much higher than the upper CI level of ``YNH-IRM.''(green bar)
These results suggest that the current YNH prediction is worse for ``Unpro'' but better for ``IRM.''
On average, GESI appears to predict YNH scores with an accuracy comparable to reported OA predictions, regardless of different HLs.


Our main concern is the outcome of open predictions in the WHIS conditions. The RMSE of ``YNH-Unpro-WHIS'' (red bar) is comparable with that of ``Unpro (open, prev. OA).''  The RMSE of  ``YNH-IRM-WHIS'' is 1\% point higher than ``IRM (open, prev. OA)'' but within the 95\% CI range, indicating no significant difference.
The RMSE of ``YNH-Dry-Unpro,'' which has no reverberation, is similar to that of ``YNH-Unpro,'' suggesting that GESI can predict SI scores regardless of reverberation.

By comparing the subjective SI scores in Fig.~\ref{fig:YNH_SbjSI_AllCondPrcs} and the predicted SI scores in Fig.~\ref{fig:YNH_PredSI_AllCondPrcs}, we can infer that the main cause of these RMSEs is the variability in subjective SI scores among YNH listeners. In this context, GESI was shown to be a reasonably accurate predictor of YNHs' SI scores.
It would be possible to further reduce RMSEs by adjusting the GESI parameters. However, since our focus here is on separating the effects of peripheral and central/cognitive factors, we will proceed to predict the SI scores of OAs using the $a_{_{YNH}}$ and $b_{_{YNH}}$ values in the next section. Before this, we will examine the predictions for each YNH listener individually.


\begin{figure}[t] 
    \centering
    \includegraphics
    [width = 1\columnwidth]
    {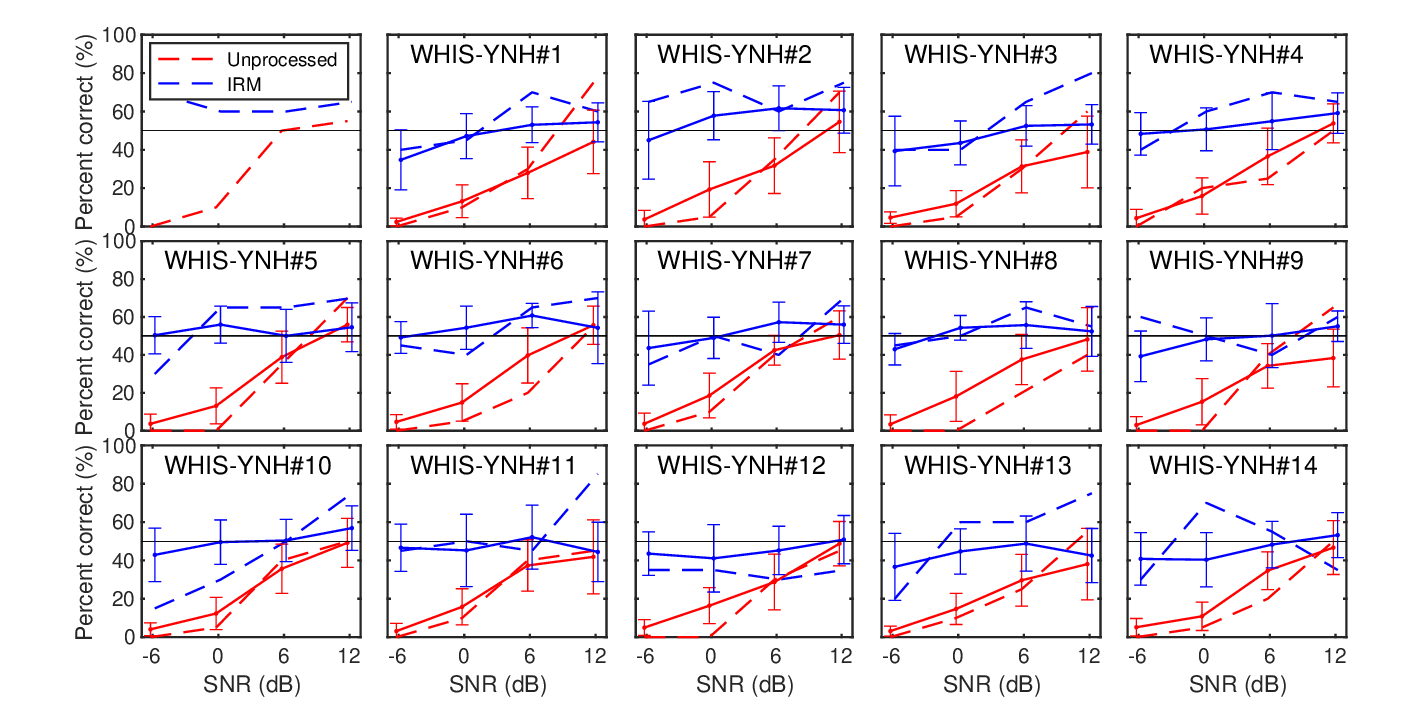}
  \caption{Subjective (dashed line) and predicted  (solid line with error bars) SI scores. The upper left panel shows the subjective SI scores of $\rm OA^{\#7}$ in the ``Unpro'' (red) and ``IRM'' (blue) conditions. The remaining panels show the SI scores of individual YNHs in the ``Unpro-WHIS'' (red) and ``IRM-WHIS'' (blue) conditions. The mean and standard deviation of the GESI predictions for the 20 words presented to listeners are shown. }
    \label{fig:YNH_PredSI_IndivLstn} 
\end{figure}

\subsection{Prediction in individual YNH}
\label{sec:PredYNH_Individual}
Figure \ref{fig:YNH_PredSI_IndivLstn} shows the SI scores of the individual YNHs in the ``Unpro-WHIS'' and ``IRM-WHIS'' conditions, which are our main concern. For comparison, the subjective SI scores of $\rm OA^{\#7}$ in the ``Unpro'' and ``IRM'' conditions are also shown in the upper left panel.
All of these predictions were made under open conditions as $a_{_{YNH}}$ and $b_{_{YNH}}$ were estimated using the ``Unpro'' condition. 
We observe a mix of cases where the subjective SI score is higher or lower than the prediction, indicating large variability. This is also evident from the variability shown in shown in Fig.~\ref{fig:CmprOA7vsYNH-WHIS}. This is likely because subjective scores reflect differences in cognitive state when listening to words. Furthermore, the 20-word lists that were randomly assigned to participants for each SNR and sound processing condition are not necessarily of the same difficulty level (see also Section \ref{sec:ExpYNH_Explanation}). This variability cannot be accounted for in the current version of GESI because it does not model cognitive processing.


\section{GESI prediction of OAs' SI scores using YNH parameters}
\label{sec:PredOA_usingYNHparam}

We examine whether the SI scores of the OAs 
could be accurately predicted by GESI using the same parameters determined for predicting the SI of the YNHs. 
If the accuracy is fairly good, it would be possible to directly compare the SI scores of the OAs with the expected SI scores based on the average characteristics of the YNHs. 


\subsection{Procedure}
\label{sec:PredOA_Procedure}

The procedure was the same as the YNH prediction described above. The hearing levels and the TMTF of each OA were entered.
The compression health parameter, $\alpha$, was set to 0.5 for all OAs, corresponding to moderate dysfunction. The other metric parameters were the same as those in the YNH prediction. An open prediction was performed for the OAs using the sigmoid parameters $a_{_{YNH}}$ and $b_{_{YNH}}$.


\begin{figure}[t]
         \centering
        \includegraphics[width=0.7\columnwidth]{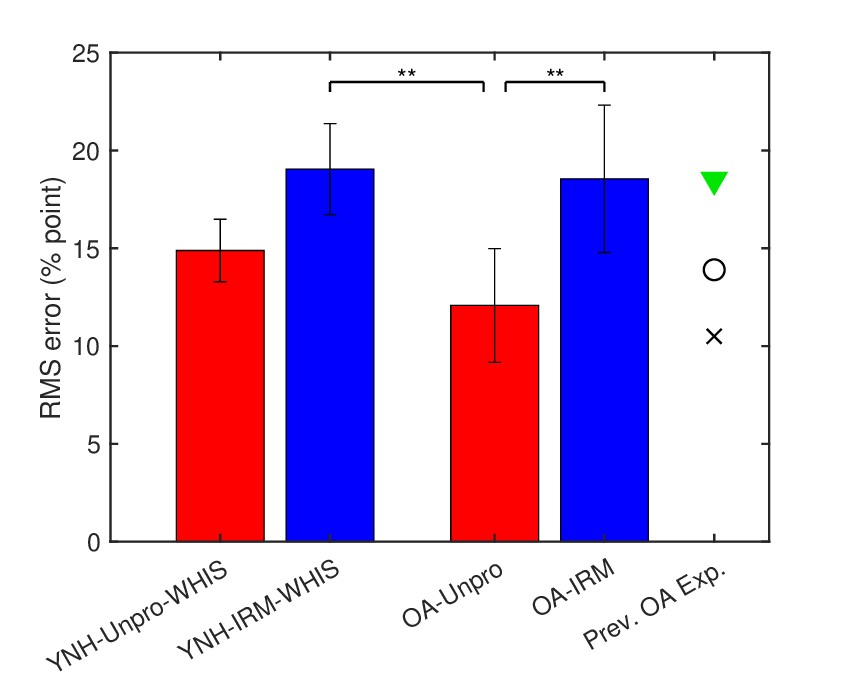} 
\caption{ Mean RMSE and 95\% CI for YNH predictions (extracted from Fig.~\ref{fig:BarRMSE_YNH-WHIS}) and OA predictions. 
Significance brackets with double asterisks (**) indicate statistically significant differences (p < 0.01) according to Tukey’s HSD multiple comparison test. The right markers show the same mean RMSEs as in Fig.~\ref{fig:BarRMSE_YNH-WHIS}. }
\label{fig:BarRMSE_YNH-WHISvsOA}
\end{figure}

\subsection{Prediction error}
\label{sec:PredOA_Error}
Figure~\ref{fig:BarRMSE_YNH-WHISvsOA} show the mean and 95\% CI for the RMSEs in the YNH and OA predictions.
The mean SI score of ``OA-Unpro'' is smaller than those of ``YNH-Unpro-WHIS'' and ``Unpro (open)'' in the previous OA study. The mean SI score of ``OA-IRM'' is almost the same as ``YNH-IRM-WHIS'' and ``IRM'' in the previous OA study. 
A two-way ANOVA of the bar graphs revealed that the main effect of the experiment (YNH vs. OA) was not significant ($F(1, 54) = 1.59, p = 0.21$), whereas the main effect of the sound processing condition (``Unpro'' vs. ``IRM'') was significant ($F(1, 54) = 16.46, p < 0.001$). The interaction between experiment and processing was not significant ($F(1, 54) = 0.78, p = 0.38$).
Post hoc multiple comparisons using Tukey’s HSD test revealed significant differences ($p < 0.01$) between  between ``OA-Unpro'' and ``YNH-IRM-WHIS,'' as well as between ``OA-Unpro'' and ``OA-IRM.'' No significant differences were observed for the remaining pairwise comparisons.
These results suggest that predicting the OAs' SI scores using the YNH parameters ($a_{YNH}$ and $b_{YNH}$) is reasonably accurate and comparable to YNH predictions.

\begin{figure}[t] 
    \centering
    \includegraphics
    [width = \columnwidth]
    {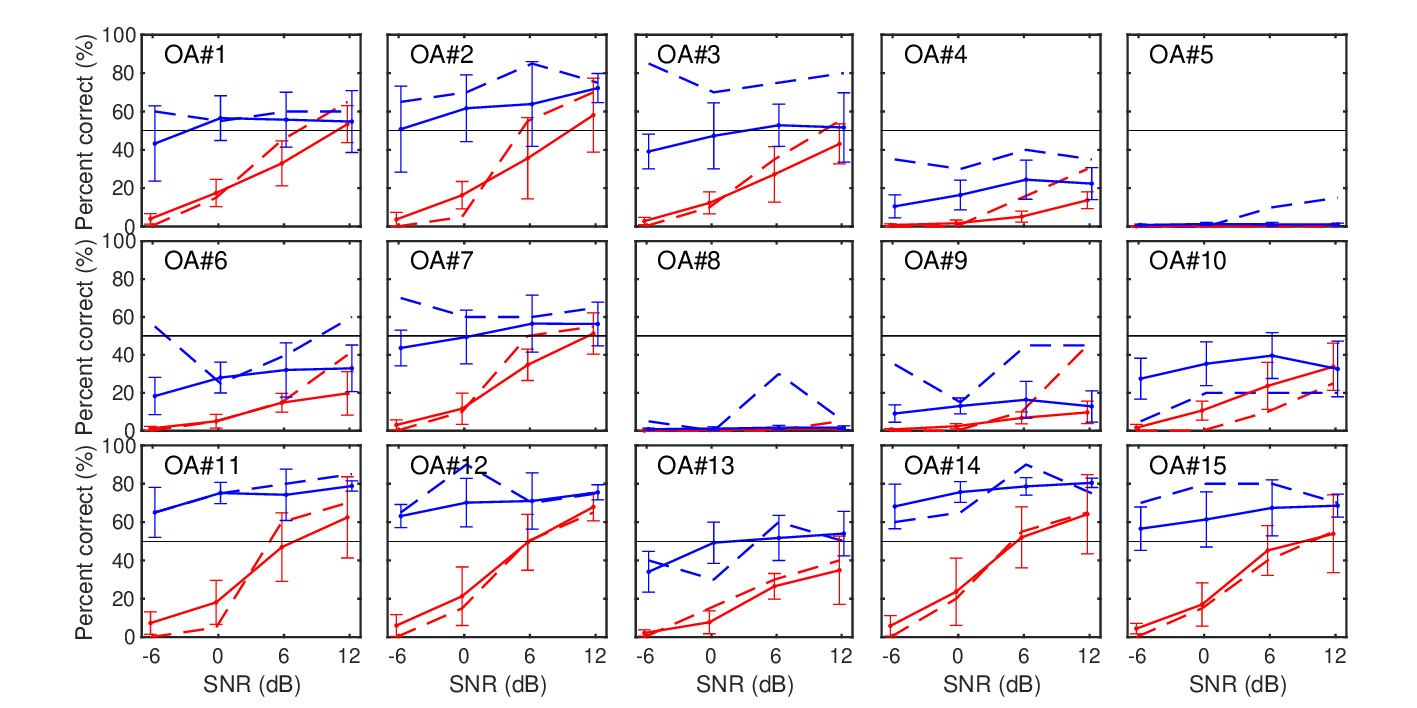}
  \caption{Subjective and predicted SI scores for individual OAs in the ``Unpro'' (red) and ``IRM'' (blue) conditions. Subjective SI was reported by \cite{yamamoto2025predicting} (dashed line). The SI scores were predicted using GESI with $a_{_{YNH}}$ and $b_{_{YNH}}$ (solid line with error bars).  The mean and standard deviation of the GESI predictions for the 20 words presented to listeners are shown. }
    \label{fig:OA_PredSI_IndivLstn} 
\end{figure}

\begin{figure}[t] 
    \centering
    \includegraphics
    [width = 1\columnwidth]
    {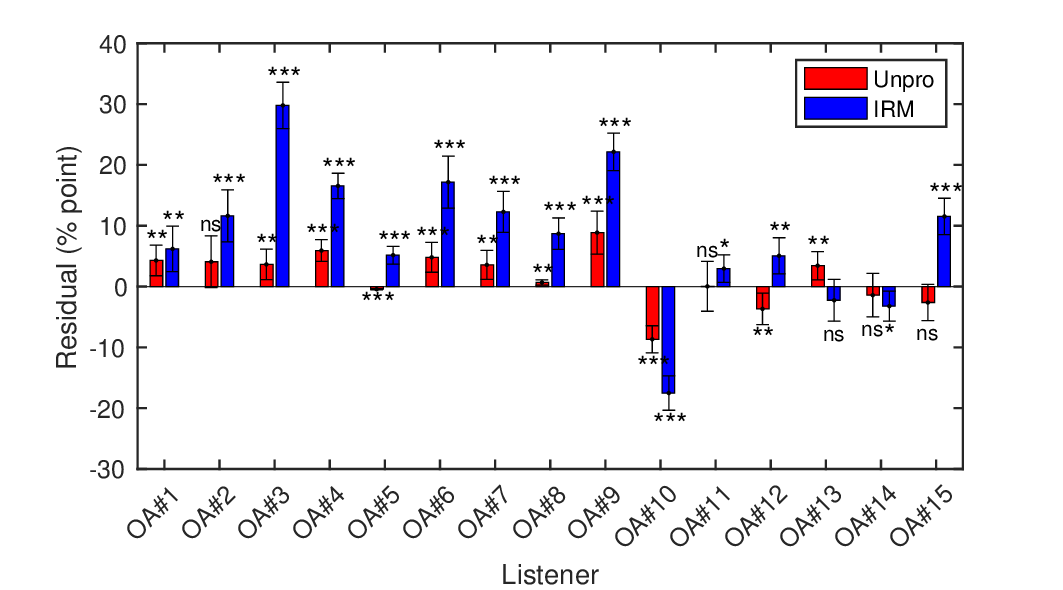}
  \caption{Residual difference between the subjective and predicted SI scores for individual OAs in the ``Unpro'' (red) and ``IRM'' (blue) conditions. 
  The bars represent the mean, and the error bars represent the 95\% CI of the difference across the four SNR conditions and 20 words per condition.
  ***: $ p <  0.001$; **: $ p <  0.01$; *: $ p <  0.05$; ns: not significant.
  }
    \label{fig:OA_DiffSbjPred} 
\end{figure}

\subsection{Prediction in individual OA}
\label{sec:PredOA_individual}


Figure~\ref{fig:OA_PredSI_IndivLstn} shows the subjective and predicted SI scores of individual OAs in the ``Unpro'' and ``IRM'' conditions. Overall, it is clear that the predictions and subjective scores generally exhibit a similar trend. These predictions are made possible by incorporating the hearing levels of each OA into GESI. Because the parameters $a_{_{YNH}}$ and $b_{_{YNH}}$ were used, the predictions reflect the average SI characteristics of the five YNHs.

In the panel of $\rm OA^{\#7}$, the predicted SI scores (solid lines with error bars) resemble the subjective SI scores for ``Unpro-WHIS'' and  ``IRM-WHIS''  in the YNH experiment, as shown in Fig.~\ref{fig:CmprOA7vsYNH-WHIS}.
Moreover, the subjective SI score at an SNR of -6~dB in ``IRM'' is much higher than predicted using the YNH parameters.
This suggests that WHIS can effectively simulate $\rm OA^{\#7}$ 's HL, and that GESI can predict the OA's SI scores reasonably well regardless of the different HLs. It seems that they can both function as an inversion system of sorts.

\subsection{Analysis of residuals between the subjective and objective SI scores}
\label{sec:PredOA_Diff}
To summarize the above results, the residual differences between the subjective and predicted SI scores across four SNRs and 20 words per condition were calculated.  Then a $t$-test was conducted for each OA and processing condition. Figure~\ref{fig:OA_DiffSbjPred} shows the results.
The residuals for many OAs were significantly greater than zero. This indicates that the subjective SI scores were higher than predicted.  The residuals in $\rm OA^{\#7}$ are also significantly positive, consistent with the WHIS-YNH experiment observation, as shown in  Fig.~\ref{fig:ExpWHISmnm24}.
In contrast, $\rm OA^{\#10}$ and a few other OAs and conditions had scores below zero. Five conditions were not significant.  
GESI predicted the SI scores using the bottom-up model with the YNH parameters $a_{_{YNH}}$ and $b_{_{YNH}}$, which reflect the average perception ability of five YNHs.
Therefore, this variability could be caused by other factors, such as variability in central and cognitive processing. This issue will be discussed in Section~\ref{sec:Discussion_Potential}.

Figure~\ref{fig:OA_DiffSbjPred} shows that the residuals are  generally greater in ``IRM'' than in ``Unpro.''  
This observation is supported by the statistical analysis of RMSE in Fig. \ref{fig:BarRMSE_YNH-WHISvsOA}. High SNR, which is achieved by the ``IRM''  speech enhancement, likely helps speech perception in OAs more effectively.

These results imply that some OAs performed better than the average YNH. This seems to contradict the commonly reported findings that speech perception ability deteriorates with age (see Section~\ref{sec:Introduction}). We need to consider why this occurs. Some thoughts on this issue will be discussed in Section~\ref{sec:Discussion}.

\subsection{Correlation analysis}
\label{sec:PredOA_Corr}

It has been demonstrated that GESI can provide reasonably accurate SI predictions even when hearing levels differ. In other words, the GESI can account for differences in hearing levels. Here, we examine the extent to which these effects persist in discrepancies between subjective and predicted SIs.

First, Pearson correlations were calculated between the average SI scores across SNRs and the PTA4, as well as the TMTF sensitivity $L_{ps}$, using data from all OAs.
The top two rows of Table~\ref{table:CorrSI_HLTMTF} show the results.
The average SI showed significantly high negative correlations with the PTA4: -0.94 ($p \ll 0.001$) for ``Unpro'' and -0.93 ($p \ll 0.001$) for ``IRM''.
This finding is consistent with the well-known fact that peripheral HL strongly correlates with SI. 
In contrast, the average SI showed a weak, non-significant correlation with $L_{ps}$.
The TMTF, as measured by broadband noise in the previous OA study, appears to have little effect on SI for these OAs.
The lack of a clear trend could be due to the dominant effect of peripheral HL, which might make the difference in TMTF less noticeable.

Then, we averaged the residuals between the predicted and subjective SIs across SNRs, and calculated the correlation between that average and either the PTA4 or $L_{ps}$.
These results are shown in the bottom two rows of Table~\ref{table:CorrSI_HLTMTF}.
The correlation coefficients for the PTA4 were less than 0.02 for both ``Unpro'' and ``IRM'', and no significant differences were observed.
Therefore, the residuals were uncorrelated with the PTA4. 
This suggests that the GESI prediction absorbed the effects of the peripheral HL. Therefore, it can be assumed that the residual reflects the effects of processes beyond those modeled by GESI.

The correlation coefficients for $L_{ps}$ were approximately -0.5, though they were not statistically significant. However, a trend toward significance was observed ($p =  0.057$ and $p =  0.064$). 
This may have become apparent because the effects of the peripheral HL was eliminated. 
Although GESI incorporates the TMTF, its effect was limited as reported by \cite{yamamoto2025predicting}. Consequently, it appears insufficient for achieving complete decorrelation, unlike in the PTA4 case. Further investigation through GESI refinement is necessary in this regard.

As a result, we were able to eliminate the effects of hearing level (PTA4) by calculating the residuals between the subjective and objective SI scores predicted by GESI with YNH parameters. This may allow us to examine other factors more accurately. We consider this to be a major contribution of this study.



\begin{table}  
\centering                 
\begin{tabular}{|c c|c c|c c|}    
\hline         
&  &\multicolumn{2}{|c|}{PTA4} &  \multicolumn{2}{|c|}{TMTF $L_{ps}$}\\                   
\hline         
Mean SI & Unpro & -0.94 &  ($p=$~1.1e-07) & -0.44 &  ($p=$~0.098) \\
\cline{2-6}      
Sbj & IRM & -0.93 & ($p=$~4.7e-07) & -0.33 & ($p=$~0.24) \\  
\hline         
Residual & Unpro & -0.0026 & ($p=$~0.99) & -0.50 &  ($p=$~0.057) \\
    \cline{2-6} 
(Sbj-Pred)   & IRM & 0.017 & ($p=$~0.95) & -0.49 & ($p=$~0.064) \\   
\hline         
\end{tabular}  
\caption{Coefficient and $p$-value of Pearson correlation, using data from all OAs. The correlations were calculated between PTA4 or TMTF $L_{ps}$ and the mean subjective SI scores (top two rows), as well as the residuals between the subjective and predicted SI scores (bottom two rows), in ``Unpro'' and ``IRM.'' }    
\label{table:CorrSI_HLTMTF}                   
\end{table}      


\section{Discussion}
\label{sec:Discussion}
\subsection{Potential reasons for better performance in OAs than in YNHs}
\label{sec:Discussion_Potential}

The subjective SI scores of many OAs were better than the predicted SI scores derived using the parameters derived in the YNH prediction, as shown in Fig.~\ref{fig:OA_DiffSbjPred}. This seems to contradict the commonly reported findings that speech perception ability deteriorates with age. There are several possible reasons for this discrepancy.

\begin{enumerate}

\item Possible bias in GESI predictions: 
These predictions here were derived using the parameters  $a_{_{YNH}}$ and $b_{_{YNH}}$, obtained from the ``Unpro'' condition of the first five YNH participants in the experimental order.
If the YNHs used to determine these parameters change, the predicted SI will naturally change as well.  However, the variability is likely less than a few percentage points. This is because the CI values for the ten repeated estimations in the previous OA study were 1.0 or less, as mentioned in Section~\ref{sec:PredYNH_Result}.
Therefore, this bias is relatively smaller than the differences observed in many OAs in Fig.~\ref{fig:OA_DiffSbjPred}.
This bias does not differ among individual OAs and appears in similar amounts in all OAs. Therefore, the relative relationships in  Fig.~\ref{fig:OA_DiffSbjPred} is expected to remain unchanged,  as will the cross-correlation coefficients in Table~\ref{table:CorrSI_HLTMTF}.

\item Incomplete modeling: It is possible that GESI, given its current settings, is incomplete. However, as described above, GESI can accurately predict the SIs of YNH and OA relative to participant variation.
As described in Section ~\ref{sec:PredOA_individual}, GESI and WHIS operate similarly to an inversion system. Therefore, the commonly used GCFB accurately reflects the hearing levels. 
However, the compression health parameter,  $\alpha$, was set to 1.0 (fully healthy) for the YNHs and 0.50 (moderately impaired) for the OAs.  As described in Appendix A of ~\cite{yamamoto2025speech}, it is currently difficult to estimate the slope of the input-output function of the cochlea with psychoacoustic tests as short as audiometry. Consequently, the $\alpha$ values of individual listeners were neither measured nor set in the previous or current study. A precise setting may affect residuals to some extent. Further study on this issue is necessary. However, since the hearing levels were accurately set in GCFB, decorrelation with PTA4 can still be achieved. 

\item Characteristics of the OA participants:
As will be explained in the next section, there is a possibility that speech perception ability in OA has actually improved or has at least been maintained.
The OA participants in the previous and current studies are healthy and have registered with the local Senior Human Resources Center to actively engage in social activities. From a young age, they are likely to have well-developed language and cognitive abilities. Engaging in communication activities with others can help prevent a decline in these abilities.

\end{enumerate}

\subsection{Improvement of speech perception in OA}
\label{sec:Discussion_Improvement}


Cognitive functions can adapt to the gradual progression of age-related HL, enabling listeners to compensate for degraded auditory input through increased reliance on higher-level linguistic and cognitive resources. Previous work has shown that OAs with HL often maintain relatively good speech understanding by using linguistic context, vocabulary knowledge, and predictive processing, despite reduced peripheral auditory sensitivity \citep{pichorafuller2015hearing, gordonsalant2020agerelated}. 
Although few studies explicitly claim that OAs are superior to YNHs, some demonstrate that OAs perform as well as or better than YNHs on tasks that rely heavily on linguistic knowledge (e.g., ~\citealp{amichetti2016multiple}).
Importantly, these abilities are not necessarily the result of explicit training but can emerge naturally through repeated exposure to speech in everyday conversational settings, reflecting experience-dependent perceptual learning and cognitive adaptation~\citep{shechtershvartzman2022speech}. 


\subsection{Decorrelation of peripheral HL and further improvement of GESI}
\label{sec:Discussion_Decorrelation}

There is no correlation between the PTA4 and the residual between the predicted and subjective SI scores. This decorrelation may be a major contribution of the study.
However, a slight correlation trend remains for the TMTF. These results suggest that GESI fairly represents the listener's peripheral auditory system, though only to a limited extent for the TMTF. As the previous OA study suggested~\citep{yamamoto2025speech}, we should improve the current version of GESI by modifying the TMTF formulation. 
Additionally, we could add modules that reflect the results of psychological experiments.
If the model achieves decorrelation, then we could consider it reasonable up to that point.
Decorrelation could be an effective indicator for modeling.

\subsection{Towards future experiments}
\label{sec:Discussion_Towards}

We found that the residual between subjective SI scores of OAs and GESI predictions using YNH parameters was uncorrelated with PTA4. 
As described in Section~\ref{sec:Introduction}, ~\cite{schoof2014role} and ~\cite{fullgrabe2015age} performed contrastive experiments between YNH and OA with NH to fact out the effects of differences in hearing levels.  However, this limitation could be overcome by using GESI. Similar experiments involving YNH and OA with HL could be conducted to more accurately evaluate the relative contributions of central and cognitive processing. Furthermore, it would also be possible to compare OAs with different hearing levels on an individual basis.



\section{Conclusion}
\label{sec:Conclusion}

We proposed a method that uses a hearing loss simulator, WHIS, and an objective SI predictor, GESI. to investigate the effects of peripheral HL and central/cognitive processing.
First, we obtained the SI scores of the 14 YNH participants using simulated HL sounds synthesized by WHIS. These sounds reflected the hearing levels of $\rm OA^{\#7}$ from a previous OA experiment ~\citep{yamamoto2025predicting}. The results showed that $\rm OA^{\#7}$ achieved higher SI scores than the average YNH listener.  Then, GESI was used to predict the SI scores in the YNH experiment. It was found that performance was comparable to that reported in the previous OA experiment.
Furthermore, to examine the SI ability of other OAs, we used GESI to predict SI scores for the 15 OAs using the sigmoid parameters $a_{{YNH}}$ and $b{_{YNH}}$ which were derived from the YNH prediction. The results showed many OAs had significantly higher subjective SI scores than predicted, while one OA had a significantly lower score. This discrepancy may reflect individual variations in central and cognitive processing, which GESI does not model.  We discussed potential reasons for better performance.  There was no correlation between PTA4 and the residuals between subjective and GESI-predicted SI scores. Thus, it seems that GESI absorbs the effects of peripheral HL. Using GESI as a strategy would provide a good framework for future studies that seek to better understand the effects of central and cognitive processing in isolation from peripheral HL.


\section*{CRediT author contribution statement}

{\bf Toshio Irino: }{Writing -- original draft, Writing -- review \& editing, 
Visualization, Validation, Software, Methodology, Investigation, Formal analysis, Conceptualization,  Supervision, Funding acquisition, Project administration.
{\bf Ayako Yamamoto: } Writing -- review \& editing, Validation, Methodology, Investigation. 
{\bf Fuki Miyazaki: }{Investigation, Data curation.}
}

\section*{Declaration of Generative AI and AI-assisted technologies in the writing process}
During the preparation of this manuscript, the authors used DeepL Write, DeepL Translate, and Microsoft Copilot to assist with improving language quality, readability, and the identification of relevant references. Following the use of these tools, the authors carefully reviewed, edited, and verified all content, and take full responsibility for the accuracy, originality, and integrity of the manuscript.

\section*{Declaration of competing interest}
The authors declare the following financial interests/personal relationships which may be considered as potential competing interests: Reports a relationship with that includes:. Has patent pending to. If there are other
authors, they declare that they have no known competing financial interests or
personal relationships that could have appeared to influence the work reported in this paper.

\section*{Data Availability Statement}
The WHIS and GESI software are available in our GitHub repository \\ (https://github.com/amlab-wakayama/).
The datasets used and/or analyzed during the current study are available from the corresponding author upon reasonable request. The source of the sound files is the database FW07, provided by the Informatics Research Data Repository at the National Institute of Informatics (NII-IDR) under a user license
(DOI: 10.32130/src.FW07). Therefore, you may need a license if you also require stimulus speech sounds.

\section*{Acknowledgments}
This research was supported by JSPS KAKENHI: Grant Numbers JP21H03468, JP21K19794,and JP24K02961. The author would like to thank Takashi Morimoto for providing the TMTF measurement software and Sota Bano and Keita Kuninaka for their assistance with the experiments.

\appendix
\renewcommand{\thetable}{\Alph{section}.\arabic{figure}}
\renewcommand{\thefigure}{\Alph{section}.\arabic{table}}

\newpage


\section{Audiogram and TMTF of YNH and OA} 
\label{sec:Appendix_AudgramTMTF}
\begin{figure}[t]
    \centering
        \begin{minipage}{0.49\textwidth}
        \centering
        \includegraphics[width=\columnwidth]{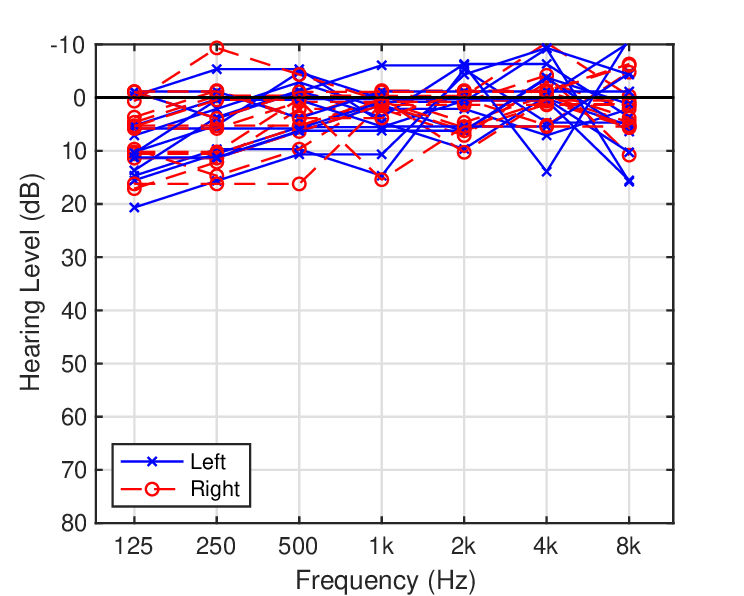}
        \subcaption{Audiograms of 14 YNHs }
        \label{fig:Audiogram_YNH}
    \end{minipage}
    \begin{minipage}{0.49\textwidth}
        \centering
        \includegraphics[width=\columnwidth]{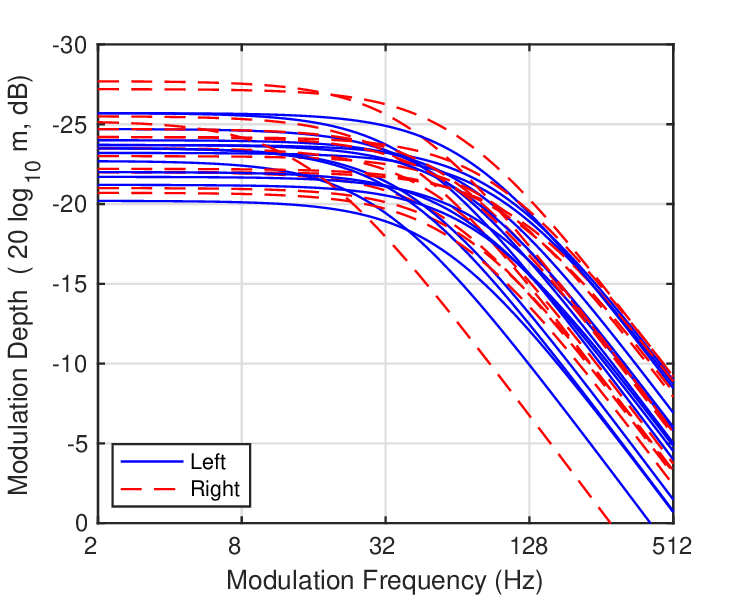} 
        \subcaption{TMTFs of 14 YNHs}
        \label{fig:TMTF_YNH}
    \end{minipage}

    \begin{minipage}{0.49\textwidth}
        \centering
        \includegraphics[width=\columnwidth]{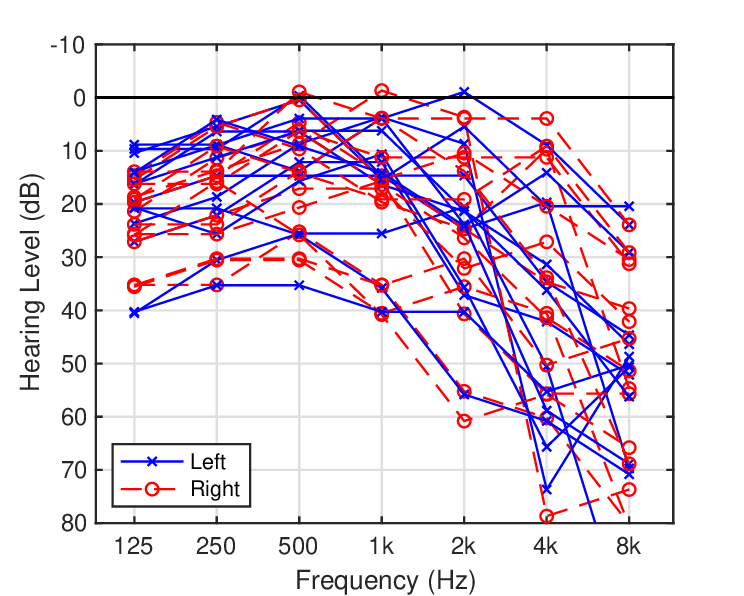}
        \subcaption{Audiograms of 15 OAs}
        \label{fig:Audiogram_OA}
    \end{minipage}
    \begin{minipage}{0.49\textwidth}
        \centering
        \includegraphics[width=\columnwidth]{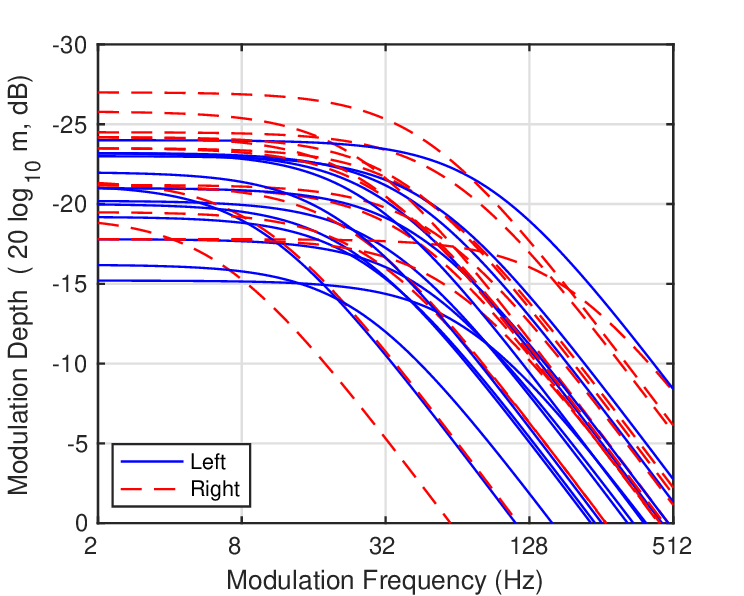} 
        \subcaption{TMTFs of 15 OAs}
        \label{fig:TMTF_OA}
    \end{minipage}
    
\caption{Audiograms and TMTFs of 14 YNHs participated in this study and 15 OAs participated in the previous OA study~\citep{yamamoto2025predicting}. Note that small random values were added to the hearing levels to distinguish the individual lines.}
\label{fig:Audiogram_TMTF_OA}
\end{figure}

The audiograms and TMTFs of the 14 YNH listeners in the current study are shown in Figs.~\ref{fig:Audiogram_YNH} and ~\ref{fig:TMTF_YNH}.
The audiograms and TMTFs of the 15 OA participants are shown in Figs.~\ref{fig:Audiogram_OA} and ~\ref{fig:TMTF_OA}.  
The OA participants were between 62 and 81 years old and their better-ear PTA4s ranged from 8.8 to 43.8 dB. Nine participants had PTA4 below 22 dB.

\section{GESI} 
\label{sec:Appendix_GESI}

For convenience, we provide an overview here to explain GESI and its parameters~\citep{yamamoto2025predicting}.
We also describe the advantages of GESI over other recent OIMs.

\begin{figure}[t] 
    \centering
     \includegraphics[scale= 0.6]{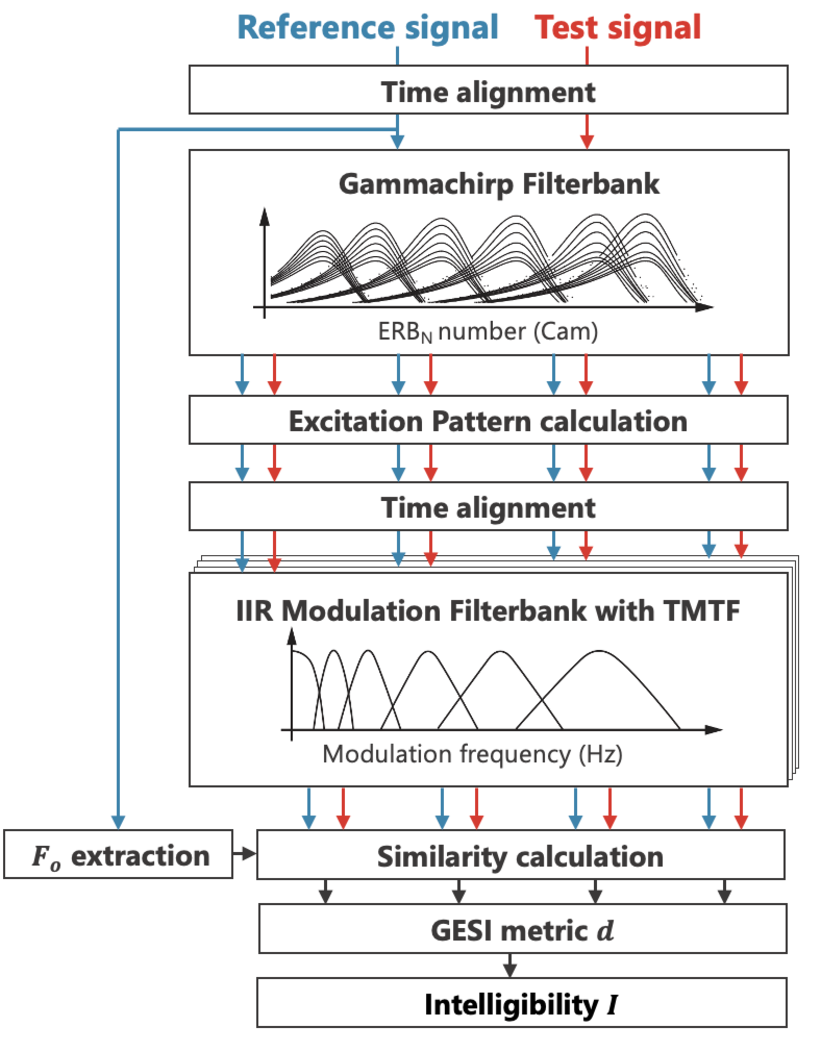}   
     \caption{Block diagram of GESI}
    \label{fig:GESI_BlockDiagram}
\end{figure}

\subsection{Algorithm of GESI} 
\label{sec:Appendix_AlgorithmGESI}
Figure \ref{fig:GESI_BlockDiagram} shows the block diagram of GESI.
The input sounds to GESI are reference ($r$) and test ($t$) signals.
First, the cross-correlation between them is computed to perform the time alignment of the speech segment. Then, both signals are analyzed with the gammachirp auditory filterbank (GCFB) \citep{irino2006dynamic,irino2023hearing}, which contains the transfer function between the sound field and the cochlea. 
The output is the Excitation Pattern (EP) sequence with 0.5~ms frame shift, something like an ``auditory'' spectrogram (hereafter referred to as EPgram).

The reference speech is always analyzed using the GCFB parameters of a typical NH listener, while the test speech is analyzed using parameters that reflect the individual listener's hearing level, that is either normal or with HL. 
This allows GESI to reflect cochlear HL resulting from dysfunction of active amplification by outer hair cells (OHCs) and passive transduction by inner hair cells (IHCs).
The required input parameters are the hearing level represented by an audiogram, and the compression health parameter ($\alpha$), which indicates the degree of health in the compressive IO function of the cochlea~\citep{irino2023hearing}. 
No dysfunction corresponds to $\alpha=1$ and completely damaged function corresponds to $\alpha=0$. 
In this study, the hearing level was set to 0~dB and $\alpha=1$ (i.e., NH level) to analyze the reference signal. The individual listener's hearing level and a default value of $\alpha=0.5$, a moderate level, were used to analyze the test signal. This is because the $\alpha$ value cannot be estimated without extensive psychoacoustic experiments (for details, see Appendix A of \citealt{yamamoto2025predicting}).

Next, the time offset between the EPgrams of the reference and test speech is compensated for each channel of the GCFB. The cross-correlation is computed to find the peak position within the maximum correction range of $\pm T_{ma}$, and the time alignment is performed accordingly.
This approach is also inspired by the strobed temporal integration mechanism of the Stabilized Wavelet--Mellin Transform (SWMT), a computational model of speech perception \citep{irino2002segregating}. Since the temporal resolution of auditory images in the SWMT is about 30 ms, we set the maximum correction range to $ T_{ma}=\pm30 \rm{ms}$.

After this correction, the EPgrams are analyzed using an infinite impulse response (IIR) version of the modulation frequency filterbank (MFB) used in GEDI ~\citep{yamamoto2018multiresolution,yamamoto2020gedi}. 
The upper limit of the modulation frequency of the MFB was restricted to 32 Hz. 
In addition, the peak gain of each filter in the MFB was set to the corresponding value of the TMTF \citep{morimoto2019twopoint} of NH and older adults obtained with the two-point method~\citep{morimoto2019twopoint}.
The peak gains of the reference ($A_j^r$) and the test ($A_j^t$) signals are defined as
%

\begin{eqnarray}
    A_j^r & = & \frac{1}{\sqrt{1 + (f_{m_j}/F_{c}^{(NH)})^{2}}} \label{eq:AjRef}\\
    A_j^t & = & \frac{10^{ (L_{ps}^{(NH)}-L_{ps}^{(HL)})/20}} 
        {\sqrt{1 + (f_{m_j}/F_{c}^{(HL)})^{2}}} \label{eq:AjTest}
    \label{eq:A_j_r_t}
\end{eqnarray}
where $j$ is the MFB channel $\{j| 2\le j \le M\}$ and $f_{m_j}$ is the MFB frequency; $L_{ps}^{(NH)}$ and $L_{ps}^{(HL)}$ are the modulation depth thresholds of the NH and HL listeners, respectively, and $F_{c}^{(NH)}$ and $F_{c}^{(HL)}$ are their cutoff frequencies. The peak gain is usually reduced in the test signal analysis since $L_{ps}^{(NH)} < L_{ps}^{(HL)}$ in many cases. Note that the first MFB filter is an LPF with a cutoff frequency of 1~Hz and we set $A_1^r = A_1^t = 1$ to maintain the DC modulation level, which is related to the AT. 
Note that this parameter setting had an insignificant effect on SI prediction in the study reported in ~\cite{yamamoto2025predicting}. Further investigation is necessary, but this setting was used in the current study.

The internal index is computed using an extended version of the cosine similarity between the MFB outputs for the reference signal ($m_{ij}^r(\tau)$) and the test signal ($m_{ij}^t(\tau)$):

\begin{eqnarray}
  S_{ij} &=& \frac{
  \sum_{\tau} w_i(\tau)\cdot m_{ij}^r(\tau)\cdot m_{ij}^t(\tau)} 
  {(\sum_{\tau}  {m_{ij}^r(\tau)}^2)^{\rho} \cdot (\sum_{\tau} {m_{ij}^t(\tau)}^2)^{\,(1-\rho)}}
  \label{eq:similarity}
\end{eqnarray}
where $i$ is the GCFB channel $\{i \,|\, 1 \le i \le N \} $, $j$ is the MFB channel $\{j \,|\, 1 \le j \le M \}$, $\tau$ is a time frame number. 
$\rho$ $\{\rho \,|\, 0 \le \rho \le 1\} $ is a weight value that allows us to handle the level difference between the reference and test sounds.
$w_i(\tau)$ is a weighting function applied to each GCFB channel. This is formulated as the product of two weighting functions, $w_i^{(SSI)}(\tau)$ and $w_i^{(Ef)}$. Thus, $w_i(\tau) = w_i^{(SSI)}(\tau) \cdot w_i^{(Ef)}$.

The weighting function $w_i^{(SSI)}(\tau)$, called SSIweight, is designed to reduce the influence of the fundamental frequency $F_o$ (e.g., gender differences) on the phonetic features. SSIweight originates from the Size-Shape Image (SSI) in the SWMT ~\citep{irino2002segregating}. The SSI is a two-dimensional image with a horizontal axis, $h$, and a vertical axis of filter channel frequency.  SSIweight has been derived for use with the EPgram.
SSIweight plays an important role in explaining experimental results on size perception from speech sounds~\citep{smith2005processing,matsui2022modelling}.
Its definition is as follows:
\begin{eqnarray} 
  w_i'(\tau) & = & \min(\frac{f_{p,i}}{h_{max}\cdot  F_o(\tau)}, 1),\nonumber \\ 
     w_i^{(SSI)}(\tau) & = & \frac{w_i'(\tau)}{\sum_{i=1}^N w_i'(\tau)},
  \label{eq:SSIweight}
\end{eqnarray}
where $f_{p,i}$ represents the peak frequency of the $i$-th channel in the GCFB. 
$F_o(\tau)$ denotes the fundamental frequency of the reference speech at frame time $\tau$, which can be estimated using tools such as the WORLD speech synthesizer \citep{morise2016world}. However, for certain sounds, such as some consonants where there is no $F_o$, a small positive value close to zero is assigned.
This ensures that the sum of $w_i^{(SSI)}(\tau)$ for all $i$ is equal to 1.
$h_{max}$ is a constant that determines the boundary between where the weight value $w_i'(\tau)$ gradually increases and where it becomes 1. This comes from the upper limit of the horizontal axis $h$ in the SSI.

The weighting function $w_i^{(Ef)}$ represents the efficiency of extracting information from the GCFB outputs above the AT. As age-related HL gradually progresses, individuals may compensate by extracting speech information from the remaining audible regions, potentially increasing overall efficiency. We formulate this effect as a weighting function.
Let \( EP_i^{(Ave)} \) be the average EP value over all time frames $\tau$ for the $i$-th GCFB channel (total $N$ channels). Channels where this value exceeds the absolute threshold (AT, i.e., 0~dB) can be considered as contributing to the information extraction. Defining the number of such audible channels as \( N_{AT} \), the weighting function \( w_i^{(Ef)} \) is formulated as follows.
\begin{equation}
    w_i^{(Ef)} = 
  \begin{cases}
     (N/N_{AT})^\eta & \text{if $EP_i^{(Ave)} >  AT $ ,} \\
     0                 & \text{otherwise}.
  \end{cases}
  \label{eq:w_i_Ef}
\end{equation}
where $\eta$ is a constant representing the efficiency.
In this study, the value of $\eta$ was set at 0.7, consistent with the previous OA study.
Efficiency may depend on cognitive factors, such as listening effort, concentration, and experience. Furthermore, efficiency may vary by channel. Future studies may be able to incorporate these factors, even partially, into GESI.

The overall similarity index $d$ is obtained by weighting and averaging $S_{ij}$ in Eq.~\ref{eq:similarity} by all $i$ and $j$ .


\begin{eqnarray}
   d & = & \frac{1}{MN} \sum_{i=1}^N \sum_{j=1}^M  w_j\, S_{ij},
  \label{eq:metric}
\end{eqnarray}
where $w_j$ is a weighting function applied to each MFB channel.
In this study, $w_j=1$ is used, but is adjustable.

The metric $d$ can be converted to word correct score (\%) or intelligibility $I$ by the sigmoid function used in STOI~\citep{taal2011algorithm} and ESTOI ~\citep{jensen2016algorithm}. 
That is: 
\begin{eqnarray}
   I & = & \frac{I_{max}}{1+\exp(a\cdot d + b)}
   \label{eq:sigmoid}
\end{eqnarray}   
where $a$ and $b$ are parameters determined from a subset of the SI scores in the experimental results using the least squares error method.
$I_{max}$ represents the maximum SI specific to the speech dataset used in the experiments. In this study, it was set to 85\% based on the SI data of the least familiar words in FW07, i.e., a subset of FW03~\citep{amano2009development}.

GESI only reflects peripheral and some central processes, not cognitive processing, until the similarity index $d$ is derived in Eq.~\ref{eq:metric}.
The sigmoid function in Eq.~\ref{eq:sigmoid} converts metric values to SI scores in a given experiment. 
Thus, only two parameters, $a$ and $b$, reflect the summary of the various, complex effects of central/cognitive processing. The SI prediction ability is restricted to what is modeled in GESI. This, in turn, contributes to separating the effects of peripheral and central/cognitive processing.


\subsection{Use of other OIMs}
There are recent OIMs that account for HL. However, using these OIMs for this study seems impractical.
HASPIv2~\citep{kates2021hearing} and HASPIw2~\citep{kates2023extending} use a small neural network (NN) to predict SI for HL listeners. More robust SI prediction methods using deep neural networks (DNNs) have recently been proposed, for example, in the Clarity Prediction Challenges ~\citep{barker20221st,barker20242nd}. However, NNs and DNNs operate as black boxes, preventing the interpretation of their reasoning behind the results. Since all peripheral and central/cognitive factors are mixed together in the prediction, it is impossible to determine the extent to which a specific factor contributes to SI. In contrast, GESI offers clear advantages over these complex models for the current purpose and is a more straightforward approach.


%

\bibliography{Reference_8Apr26}

\end{document}